\documentstyle[11pt]{article}
\newcommand{\beq}{\begin{equation}}
\newcommand{\eeq}{\end{equation}} 

\newcommand{\beqa}{\begin{eqnarray}}
\newcommand{\eeqa}{\end{eqnarray}}

\newcommand{\K}{{\cal K}}

\def\half{\frac{1}{2}}
\def\opone{\leavevmode\hbox{\small1\kern-3.8pt\normalsize1}}
\def\ua{\uparrow}  
\def\da{\downarrow}  
 
\begin{document}

\title{Non-Markovian Quantum State Diffusion}
\author
{L. Di\'osi$^1$, N. Gisin$^2$ and W. T. Strunz$^3$\\
\protect\small\em $^1$Research Institute for Particle and Nuclear Physics,
1525 Budapest 114, POB 49, Hungary \\
\protect\small\em $^2$Group of Applied Physics, University of Geneva, 
1211 Geneva 4, Switzerland \\
\protect\small\em $^3$Fachbereich Physik, Universit\"at GH Essen, 
45117 Essen, Germany}
\date{\today}

\maketitle

\begin{abstract}
A nonlinear stochastic Schr\"odinger equation for pure states describing 
non-Markovian diffusion of quantum trajectories and
compatible with non-Markovian master equations is presented.
This provides an unravelling of the evolution of any quantum system 
coupled to a finite or infinite number of harmonic
oscillators without any approximation. Its power is illustrated by several 
examples, including 
measurement-like situations, dissipation, and quantum Brownian motion.
Some examples treat this environment phenomenologically as an infinite 
reservoir with fluctuations of arbitrary correlation. In other examples 
the environment consists of a finite number of oscillators. 
In such a quasi-periodic case we show the reversible decay of a 
'Schr\"odinger cat' state.
Finally, our description of open systems is compatible with different 
positions of the 'Heisenberg cut' between system and environment.
\end{abstract}

\section{Introduction}\label{introduction}
In quantum mechanics, a mixed state, represented by a density matrix
$\rho_t$, describes both an
ensemble of pure states and the (reduced) state of a system entangled 
with some other system, here consistently called 'the
environment'. In both cases the time evolution of 
$\rho_t$ is given by a linear map
\beq\label{rhotevol}
\rho_t={\cal L}_t\rho_0,
\eeq
which describes the generally non-Markovian evolution of the system under 
consideration. Such equations describe both an open system in
interaction with infinite reservoirs with an arbitrary correlation, or
a system entangled with a finite environment. 
In almost all cases, the general eq. (\ref{rhotevol})
cannot be solved analytically. Even numerical simulation is most often 
beyond today's algorithms and computer capacities, and thus, the solution
of (\ref{rhotevol}) remains a challenge. 

In the Markov limit, eq. (\ref{rhotevol})
simplifies and reduces to a master
equation of Lindblad form \cite{Lindbl76}
\beq\label{Lindblad}
\frac{d}{dt}\rho_t= -i[H,\rho_t] +\frac{1}{2}
\sum_m\left([L_m\rho_t, L_m^{\dagger}] + [L_m,\rho_t L_m^{\dagger}]
\right),
\eeq
where $H$ is the system's Hamiltonian and the
operators $L_m$ describe the effect of 
the environment in the Markov approximation. This approximation is often
very useful because it is
valid for many physically relevant situations and because analytical or
numerical solutions can be found.

In recent years, a breakthrough in solving the Markovian master equation 
(\ref{Lindblad}) has been made through the discovery of {\it stochastic 
unravellings} of the density operator dynamics.
An unravelling is a stochastic Schr\"odinger equation 
for states $|\psi_t(z)\rangle$, driven by a certain noise $z_t$ 
such that the mean of the solutions of the stochastic equation 
equals the density operator 
\beq\label{rhopsi}
\rho_t = M\Bigl[|\psi_t(z)\rangle\langle\psi_t(z)|\Bigr].
\eeq
Here $M[\ldots]$ denotes the ensemble mean value over the classical
noise $z_t$ according to a certain distribution functional $P(z)$. 

The simplest stochastic Schr\"odinger equations unravelling the density
matrix evolution
are linear and do not preserve the norm of $\psi_t(z)$. Such an 
unravelling is merely a mathematical relation.
To be truly useful, one should derive 
unravellings in terms of the corresponding normalized states
\beq\label{normpsi}
\tilde\psi_t(z)=\frac{\psi_t(z)}{\Vert\psi_t(z)\Vert}
\eeq
where now relation (\ref{rhopsi}) can be interpreted as
an unravelling of the mixed state $\rho_t$ into an ensemble of pure 
states.  Of course, using the normalized states $\tilde\psi_t(z)$ 
requires a change of the distribution $P(z)\rightarrow \tilde P_t(z)$
in order to ensure the correct ensemble mean, with
\beq\label{Girs}
\tilde P_t(z)\equiv\Vert\psi_t(z)\Vert^2 P(z)
\eeq
so that the eq.(\ref{rhopsi}) remains valid for the normalized solutions,
\beq\label{rhonormpsi}
\rho_t = \tilde M_t\Bigl[|\tilde\psi_t(z)\rangle\langle\tilde\psi_t(z)
|\Bigr].
\eeq

We refer to this change (\ref{Girs}) of the probability measure as a
Girsanov transformation \cite{GatGis91} -
other authors refer to "cooking the probability"
or to "raw and physical ensembles" \cite{Ghirar90a},
or to "a priori and a posteriori states" \cite{BarBel91}.

In the case of Markovian master equations of Lindblad form 
(\ref{Lindblad}), several such unravellings (\ref{rhonormpsi}) are known.
Some unravellings involve jumps at random times, others have continuous
solutions.
The Monte-Carlo wave function method \cite{Dalibard}, 
sometimes called quantum jump trajectories \cite{Carmichael,PleKni97}, 
is the best known example of the first class, whereas
the Quantum State Diffusion (QSD) unravelling \cite{QSD3} is typical of 
the second class.
All these unravellings have been used extensively over recent years, as 
they provide
useful insight into the dynamics of continuously monitored (individual)
quantum processes \cite{CohZam93,Spiller94}. In
addition, they
provide an efficient tool for the numerical solution of master equations. 
It is thus desirable to extend the powerful concept of stochastic
unravellings to the more general case of non-Markovian evolution. 
First attempts towards this goal using linear equations can be found in 
\cite{Pearle96}, other 
authors have tackled this problem by adding fictious
modes to the system in such a way as to make the enlarged, hypothetical 
system's dynamics Markovian again \cite{Imamog94,Garraw97,Bay97}.
In our approach, by contrast, the system remains as small as possible
and thus the corresponding stochastic Schr\"odinger equation becomes 
genuinely non-Markovian.

Throughout this paper we assume a normalized initial state 
$\psi_0(z)\equiv\psi_0$ 
of the system, independent of the noise at $t=0$.
Such a choice corresponds to a pure initial
state $\rho_0=\vert\psi_0\rangle\langle\psi_0\vert$ for the quantum 
ensemble and correspondingly, to a factorized initial state 
$\rho_{tot} = \rho_0\otimes\rho_{env}$ of the total density operator of
system and environment.
 
In this article we present for the first time the nonlinear non-Markovian
stochastic Schr\"odinger equation that unravels the dynamics of a system
interacting with an arbitrary 'environment' of harmonic oscillators, 
finite or infinite in number. For a brief overview of the underlying
microscopic model see Appendix C. In the Markov limit, this unravelling
reduces to QSD \cite{QSD3} and will therefore be referred to as 
{\it non-Markovian
quantum state diffusion}. Our results are based on the linear theories
presented in \cite{Diosi96,Strunz96b}, where the problem of
non-Markovian unravellings was tackled from two quite different approaches. 
The linear version of the non-Markovian stochastic Schr\"odinger equation 
relevant for this paper, unifying these first attempts,
was presented in \cite{DioStr97} for unnormalized states. 

Here we present examples of the corresponding normalized and thus more 
relevant theory. We include cases where the environment is 
treated phenomenologically, represented by an
exponentially decaying bath correlation function, and cases
where the 'environment' consists of only a finite, small
number of oscillators - in section \ref{cats} of even just a single 
oscillator.
The latter case corresponds to periodic (or quasi periodic) systems, 
that is to extreme non-Markovian situations.
Before presenting examples in sections \ref{ex}, \ref{moreex}, and 
\ref{cats}, all the 
basic equations are summarized in the next section \ref{eqs}.
Several open problems are discussed in section \ref{prob}, while the
conclusive section \ref{concl} summarizes the main achievements.

\section{Basic equations}\label{eqs}
In this section we summarize all the basic equations. Let us start by
recalling the case of Markovian QSD, providing an unravelling of the
Markovian Lindblad master equation (\ref{Lindblad}).
\subsection{Markov case}\label{MarkovSection}
The linear QSD equation for unnormalized states reads
\beq\label{MLQSD}
\frac{d}{dt}\psi_t = -iH\psi_t + L\psi_t\circ z_t - \frac{1}{2}
L^\dagger L \psi_t ,
\label{dpsiMarkov}
\eeq
where $z_t$ is a white complex-valued Wiener process of zero mean and
correlations
\beq\label{Mcorrelation}
M[z_t^*z_s]=\delta(t-s),\;\;M[z_t z_s]=0,
\eeq
and $\circ$ denotes the Stratonovich product \cite{HasEza80}. 

The solutions of  eq.(\ref{MLQSD}) unravel the density matrix
evolution according to the master eq.
(\ref{Lindblad}) through the general relation
(\ref{rhopsi}). Here, equation (\ref{MLQSD}) is written for a single 
Lindblad operator $L$, but it can be straightforwardly
generalized by including a sum over all Lindblad operators $L_m$, each 
with an independent complex Wiener process $z_m$. 

The simple linear equation (\ref{MLQSD}) has two drawbacks. 
First, its physical interpretation is
unclear because unnormalized state vectors do not represent pure states. Next, 
its relevance
for numerical simulation is severely reduced by the fact that the norm 
$\Vert\psi_t(z)\Vert$ 
of the solutions
tends to 0 with probability 1 (and to infinity with
probability 0, so that
the mean square norm is constant). Hence, in practically all numerical 
simulations of (\ref{MLQSD}) the  
norm tends to 0, while the contribution to the density matrix in
(\ref{rhopsi}) is dominated 
by very rare realizations of the noise $z$.

Introducing the normalized states (\ref{normpsi}) removes both these 
drawbacks.
As a consequence, the linear eq.(\ref{MLQSD}) is transformed into a 
nonlinear equation for $\tilde\psi_t(z)$. 
In this Markov case, the result of Girsanov transforming the noise
according to (\ref{Girs})
and normalizing the state is well known \cite{GatGis91,Ghirar90a}, 
it is the following QSD
evolution equation for the normalized states
\beq\label{MQSD}
\frac{d}{dt}
\tilde\psi_t = -iH\tilde\psi_t  + (L-\langle L\rangle_t)\tilde\psi_t\circ 
(z_t + \langle L^\dagger\rangle_t)
 -\frac{1}{2} (L^\dagger L-\langle L^\dagger L\rangle_t)\tilde\psi_t,
\eeq
where $\langle L \rangle = \langle\tilde\psi_t|L|\tilde\psi_t\rangle$.
This equation is the standard QSD equation for the Markov case
written as a Stratonovich
stochastic equation. Notice that it appears in its It\^o version in
references \cite{QSD3}. The effect of the Girsanov transformation is the
appearance of the shifted noise 
\beq\label{Mshift}
z_t + \langle L^\dagger\rangle_t,
\eeq
entering (\ref{MQSD}),
where $z_t$ is the original process of (\ref{MLQSD}).
The effect of the normalization is the subtraction of the operator's 
expectation values.

\subsection{Non-Markovian case}
In the non-Markovian case, the linear stochastic Schr\"odinger equation
generalizing (\ref{MLQSD}) was derived 
in reference \cite{DioStr97}, it reads
\beq\label{LQSD}
\frac{d}{dt}\psi_t=-iH\psi_t + L\psi_t z_t - 
L^\dagger\int_0^t \alpha(t,s)\frac{\delta\psi_t}{\delta z_s}ds.
\eeq
It unravels the reduced dynamics of a system coupled to an
arbitrary 'environment' of harmonic oscillators - see
Appendix C for a brief overview. Thus, eq.(\ref{LQSD}) represents an
unravelling of a certain (standard) class of general non-Markovian
reduced dynamics as in (\ref{rhotevol}).
The structure of eq.(\ref{LQSD}) is very similar to the Markovian linear
equation (\ref{MLQSD}): the isolated system dynamics is 
Schr\"odinger's equation with some Hamiltonian $H$. The stochastic 
influence of the environment is described by a
complex Gaussian process $z_t$ driving the system through the Lindblad
operator $L$. While this is a white noise
process in the Markov case, here it is a colored process with zero 
mean and correlations
\beq\label{correlation}
M\left[z_t^*z_s\right]=\alpha(t,s),\;\;\; M\left[z_t z_s\right]=0,
\eeq
where the Hermitian $\alpha(t,s) = \alpha^*(s,t)$ is the environment
correlation function. Its microscopic expression can be found in 
Appendix C. In this paper, we sometimes but not always 
adopt a phenomenological point
of view and will often choose $\alpha(t,s)$ to be an 
exponential ($\frac{\gamma}{2}\exp(-\gamma|t-s|-i\Omega(t-s)$), decaying on
a finite environmental 'memory' time scale $\gamma^{-1}$, and oscillating
with some environmental central frequency $\Omega$. 
The Markov case emerges in the limit $\gamma \rightarrow \infty$. 
In the most extreme
non-Markovian case, when the `environment' consists of just a single
oscillator of frequency $\Omega$, we have the periodic 
$\alpha(t,s)= \exp(-i\Omega(t-s))$. 
Finally, 
the last term of eq.(\ref{LQSD}) is the non-Markovian generalization 
of the last term of the Markovian linear QSD eq. (\ref{MLQSD}). This 
term is highly non-trivial and reflects the origin of the difficulties 
of non-Markovian unravellings.

One can motivate equation (\ref{LQSD}) on several grounds. First, it was 
originally derived from a microscopic system-environment model 
\cite{DioStr97}. In the original derivation the correlation function 
$\alpha(t,s)$ describes the correlations of environment 
oscillators with positive frequencies. However, as can be seen
in Appendix C, any positive definite $\alpha(t,s)$ can formally be 
obtained from some suitably chosen environment that possibly includes 
negative frequency oscillators (Hamiltonian not bounded from below). 

Next, as a second motivation, we sketch a direct proof that 
eq. (\ref{LQSD}) defines an evolution equation (\ref{rhotevol}) for 
density operators.  This ensures that the stochastic equation is
compatible with the standard description of mixed quantum states 
\cite{Gisin89,GisRig95}.
Let $\rho_0=\sum_j p_j |\psi^{(j)}_0\rangle\langle\psi^{(j)}_0|$
be any decomposition of the density operator
at the initial time $0$ (recall that at time zero the system and
environment are assumed uncorrelated). What needs to be proven is that
$\rho_t$ is a function of $\rho_0$ only, where 
$\rho_t\equiv\sum_j p_j M[|\psi^{(j)}_t\rangle\langle\psi^{(j)}_t|]$.
This guarantees that $\rho_t$ does not depend on the
decomposition of $\rho_0$ into a mixture of pure states 
$\{|\psi^{(j)}_0\rangle\}$. For this purpose we notice that
the solution $\psi_t$ of (\ref{LQSD}) is analytic in $z$ and is 
thus independent of $z^*$. Hence we find
$\frac{\delta|\psi_t\rangle}{\delta z_s}\langle\psi_t| 
=\frac{\delta(|\psi_t\rangle\langle\psi_t|)}{\delta z_s}$.
Accordingly, the evolution equation of
$|\psi_t\rangle\langle\psi_t|$ is linear: 
it depends linearly on $|\psi_0\rangle\langle\psi_0|$.
Since the mean
$M$ is also a linear operation, $\rho_t$ depends linearly on $\rho_0$. 
Finally, the positivity of $\rho_t$ is guaranteed by the existence of a 
pure state decomposition and its normalization follows from the fact 
that eq. (\ref{LQSD}) preserves the norm in the mean, 
$M[\Vert\psi_t\Vert^2]=const$ as shown in Appendix B.

Thirdly, another set of motivations for eq. (\ref{LQSD}) is
provided by the numerous examples of the next sections of this article 
and by the fact that, by full analogy with the Markov case, there 
exists a corresponding non-linear equation for normalized states, as will 
be shown in the remainder of this section.

To summarize, eq.(\ref{LQSD}) is the basic equation for non-Markovian 
linear QSD. The functional derivative under
the integral indicates that the evolution of the state $\psi_t$ at time
$t$ is influenced by
its dependence on the noise $z_s$ at earlier times s. 
Admittedly, this functional derivative is the cause for the difficulty
of finding solutions of eq (\ref{LQSD}) in the general case, 
even numerical solutions.

We tackle this problem by noting that
the linear equation (\ref{LQSD}) may be simplified with the Ansatz
\beq\label{Ansatz}
\frac{\delta\psi_t}{\delta z_s}= \hat O(t,s,z)\psi_t,
\eeq
where the time and noise dependence of the operator $\hat O(t,s,z)$ can 
be determined from the consistency condition
\beq\label{consistentAnsatz}
\frac{d}{dt}\frac{\delta\psi_t}{\delta z_s} = \frac{\delta}{\delta
z_s}\dot\psi_t
\eeq
with the linear equation (\ref{LQSD}). The Ansatz (\ref{Ansatz}) is
completely general and hence, once the operator $\hat O(t,s,z)$ is known,
the linear non-Markovian QSD equation (\ref{LQSD})
takes the more appealing form
\beq\label{sLQSD}
\frac{d}{d t}\psi_t=-iH\psi_t + L\psi_t z_t - 
L^\dagger\int_0^t \alpha(t,s) \hat O(t,s,z)ds \psi_t.
\eeq
We are going to show in the subsequent 
sections how to determine $\hat O(t,s,z)$ for many interesting and 
physically relevant examples. In most of these cases, in fact, the 
operator $\hat O$ turns out to be independent of the noise $z$ 
and takes a simple form.

Being the non-Markovian generalization, eq.(\ref{LQSD}) or equivalently 
eq.(\ref{sLQSD}) suffers from the 
same drawbacks as its Markov limit (\ref{MLQSD}): the norm of its 
solutions tend to 0 with
probability 1. And the cure will be similar. One introduces the 
normalized states (\ref{normpsi}) and substitutes the linear stochastic 
Schr\"odinger equation (\ref{sLQSD}) by the corresponding 
non-linear one. Its explicit form can be rather involved as will be 
demonstrated in the following sections. 

The derivation of the desired evolution equation of the normalized states 
$\tilde\psi_t$ requires two steps: Taking into account the Girsanov 
transformation of the noise (\ref{Girs})
and normalization.  In Appendix B we prove that 
the non-Markovian Girsanov transformation for the noise probability 
distribution $\tilde P_t(z)$ (\ref{Girs})
corresponds to a time dependent shift
of the original process $z_t$,
\beq\label{shift}
\tilde z_t = z_t + \int_0^t \alpha(t,s)^* \langle L^\dagger\rangle_s ds.
\eeq 

This shift and the normalization of the 
state $\psi_t$ results, as shown in Appendix B, 
in the non-linear, non-Markovian QSD equation for the normalized 
state vectors
$\tilde\psi_t$, which takes the ultimate form
\beqa\label{NMQSD}
\frac{d}{d t}\tilde\psi_t 
& = & -iH\tilde\psi_t + \left(L-\langle L\rangle_t\right)\tilde\psi_t 
\tilde z_t \\
\nonumber
& & - \int_0^t\alpha(t,s)
\left((L^\dagger - \langle L^\dagger \rangle_t)\hat O(t,s,\tilde z) 
-\langle (L^\dagger - \langle L^\dagger \rangle_t) \hat O(t,s,\tilde z)
\rangle_t\right)ds
\;\tilde\psi_t,
\eeqa
where $\tilde z_t$ is the shifted noise (\ref{shift}).      

Equation (\ref{NMQSD}) is the central result of this paper, the
non-Markovian, normalized stochastic Schr\"odinger equation that
unravels the reduced dynamics of a system in interaction with an
arbitrary 'environment' of harmonic oscillators - encoded by the
properties of the environment correlation function $\alpha(t,s)$.
In the following sections we will give many interesting examples 
of this {\it non-Markovian Quantum State Diffusion} 
equation (\ref{NMQSD}).

\section{Spin $\half$ examples}\label{ex}
In this section we use spin $\half$ examples to illustrate general methods 
to solve the non-Markovian QSD equations (\ref{LQSD}) 
(or (\ref{sLQSD})) and (\ref{NMQSD})
respectively. These are generally numerical, though sometimes analytical,
solutions
which illustrate certain new features of non-Markovian QSD, unknown
in the Markov theory. 
Throughout this section $\vec\sigma$ denote the Pauli matrices.

\subsection{Measurement-like interaction}\label{sigma_z}
This is the simplest example, hence we present it in some detail.
Let $H=\frac{\omega}{2}\sigma_z$, $L=\lambda\sigma_z$ with $\lambda$ a real
number parameterizing the strength of the interaction. The harmonic 
oscillator environment is encoded by its correlation function 
$\alpha(t,s)$ which is
left arbitrary in this section.  First, in order to eliminate the
functional derivative in (\ref{LQSD}), we assume as an Ansatz
\beq\label{spinansatz}
\frac{\delta\psi_t}{\delta z_s}=\lambda\sigma_z\psi_t,
\eeq
i.e. we choose $ \hat O(t,s,z) = \lambda \sigma_z$ independent of $t,s$ 
and $z$ in (\ref{Ansatz}).
It is straightforward to show that, indeed, this Ansatz is compatible
with (\ref{consistentAnsatz}), i.e. it solves the fundamental linear
equation (\ref{LQSD}).

The corresponding non-linear, non-Markovian QSD eq.(\ref{NMQSD})
for the normalized state $\tilde\psi_t$ reads
\beq\label{NMQSDSPIN}
\frac{d}{dt}\tilde\psi_t = -i\frac{\omega}{2}\sigma_z\tilde\psi_t + 
\lambda(\sigma_z - \langle\sigma_z\rangle_t)\tilde\psi_t
\left(z_t+\lambda\int_0^t\alpha(t,s)^*\langle\sigma_z\rangle_s ds
+\lambda\int_0^t\alpha(t,s)ds\;\;\langle\sigma_z\rangle_t\right).
\eeq
This equation is the generalization of the Markov QSD equation 
(\ref{MQSD}) for general environment correlations $\alpha(t,s)$.
Notice that, indeed, (\ref{NMQSDSPIN}) reduces to the corresponding 
Markov QSD equation (\ref{MQSD}) 
in the limit of a delta-correlated environment 
(one has $\int_0^t\alpha(t,s)f(s)ds\rightarrow\half f(t)$ for any 
function $f(t)$).

Eq. (\ref{NMQSDSPIN}) shows the effect of the non-Markovian Girsanov 
transformation (\ref{Girs}). It induces not only the shifted 
noise 
(\ref{shift}), but also leads to an additional shift due to 
the implicit $z_t$-dependence of $\tilde\psi_t$, as explained in detail
in Appendix B. Numerical simulations of (\ref{NMQSDSPIN}) are shown
below.

In order to find the reduced density matrix of this model, we solve 
analytically the linear non-Markovian QSD equation (\ref{sLQSD}).
Using (\ref{spinansatz}) we find
\beq\label{LQSDSPIN}
\frac{d}{dt}\psi_t = -i\frac{\omega}{2}\sigma_z\psi_t + 
\lambda\sigma_z\psi_t z_t + \lambda^2\int_0^t\alpha(t,s)^*ds\psi_t.
\eeq
From the explicit solution of this equation we obtain the expression
for the ensemble mean
\beq\label{rhotSPIN}
\rho(t) \equiv M\left[|\psi_t\rangle\langle\psi_t|\right]=
\pmatrix{\rho_{11}(0) & \rho_{12}(0)e^{-F(t)}\cr \rho_{21}(0)e^{-F(t)^*} 
& \rho_{22}(0)},
\eeq 
with $F(t) = i\omega t + 2\lambda^2\int_0^t ds\int_0^s
du(\alpha(s,u)+\alpha^*(s,u))$.
Taking the time derivative, one can show that this density matrix is 
the solution of the following non-Markovian master equations
\beqa\label{drhoSPIN1}
\dot\rho_t&=&-i\frac{\omega}{2}[\sigma_z,\rho_t]
- \frac{\lambda^2}{2}\int_0^t(\alpha(t,s)+\alpha^*(t,s))ds~
[\sigma_z,[\sigma_z ,\rho_t]] \\ 
\label{drhoSPIN2}
&=&-i\frac{\omega}{2}[\sigma_z,\rho_t] +
\int_0^t\K(t,s)\rho_s ds,
\eeqa
where the "memory super-operator" $\K(t,s)$ acts as follows on any 
operator $A$:
\beq
\K(t,s) A = -\frac{\lambda^2}{2}(\alpha(t,s)+\alpha(t,s)^*)
e^{-2\lambda^2\int_s^tdu\int_0^udv(\alpha(u,v)+\alpha(u,v)^*)}
[\sigma_z,[\sigma_z, A]].
\eeq

Let us now turn to actual simulations of this example.
In Figure 1a we show non-Markovian QSD trajectories from solving
(\ref{NMQSDSPIN}) numerically with $\lambda^2=2\omega$ and an 
exponentially decaying environment correlation function 
$\alpha(t,s) = \frac{\gamma}{2}\exp(-\gamma|t-s|)$ with $\gamma=\omega$
(solid lines).
For this exponentially decaying environment correlation function
the asymptotical solution is either 
the up state or the down state ($\langle\sigma_z\rangle = \pm 1$), 
while the ensemble mean 
$M[\langle\sigma_z\rangle]$ remains constant (dashed line). 
Thus, as in the standard Markov QSD case, the two outcomes
`up' or `down' appear with the expected quantum probability: 
Prob(${\lim\atop{t\rightarrow\infty}}\psi_t=|\uparrow\rangle$)
=$|\langle\uparrow|\psi_0\rangle|^2$.
Notice that for these non-Markovian situations, the quantum
trajectories are far smoother than their white-noise counterparts
of Markov QSD \cite{QSD3}. We emphasize that if the environment
consists of only a finite number of oscillators, represented by a
quasi-periodic correlation function $\alpha(t,s)$, no such
reduction to an eigenstate will occur.

In Figure 1b we
compare the average over $10000$ trajectories of the non-Markovian QSD
equation (\ref{NMQSD}) with the analytical ensemble mean (\ref{rhotSPIN})
and see very good agreement.
This confirms that indeed,
both the memory integrals in eq.(\ref{NMQSDSPIN}) arising from
the Girsanov transformation of the noise are needed to ensure the 
correct ensemble mean.  

\subsection{Dissipative interaction}\label{DampedSpin}
This is the simplest example with a non-selfadjoint Lindblad operator.
Again we set $H=\frac{\omega}{2}\sigma_z$, but now we choose
$L=\lambda\sigma_-\equiv\lambda\half(\sigma_x-i\sigma_y)$ describing spin
relaxation. Also, the environmental correlation function $\alpha(t,s)$ 
and thus the quantum harmonic oscillator environment
can be chosen arbitrary.

First we have to replace the functional derivative in (\ref{LQSD}), and
we try an Ansatz (\ref{Ansatz}) of the form
\beq \label{AnsatzDamped}
\frac{\delta\psi_t}{\delta z_s}=f(t,s)\sigma_-\psi_t,
\eeq
with $f(t,s)$ a function to be determined. 
The consistency condition (\ref{consistentAnsatz}) of our Ansatz 
(\ref{AnsatzDamped}) leads to the condition on $f(t,s)$:
\beqa
\partial_t f(t,s)\sigma_-\psi_t&=&[-i\frac{\omega}{2}\sigma_z-\lambda
F(t)\sigma_+\sigma_-, f(t,s)\sigma_-]\psi_t \\
&=&(i\omega+\lambda F(t))f(t,s)\sigma_-\psi_t
\eeqa
with 
\beqa\label{Fsigmadamped}
F(t)\equiv\int_0^t\alpha(t,s)f(t,s)ds.
\eeqa

Hence, if $\sigma_-\psi_t\ne 0$, the function $f(t,s)$ must satisfy the
following eq.:
\beq
\partial_t f(t,s)=(i\omega+\lambda F(t))f(t,s)
\label{dotfDamped}
\eeq
with initial condition $f(s,s)=\lambda$. 
The corresponding non-Markovian QSD equation (\ref{NMQSD}) for 
normalized state vectors $\tilde\psi_t$ reads
\beqa\label{NMQSDsigmadamped}
\dot{\tilde\psi}_t&=&-i\frac{\omega}{2}\sigma_z\tilde\psi_t-\lambda
F(t)(\sigma_+\sigma_--\langle\sigma_+\sigma_-\rangle_t)\tilde\psi_t \\ 
\nonumber
&+&\lambda(\sigma_--\langle\sigma_-\rangle_t)\tilde\psi_t
\left(z_t+\lambda\int_0^t\alpha(t,s)^*\langle\sigma_+\rangle_s ds
+\langle\sigma_+\rangle_t F(t)\right)
\eeqa
with $F(t)$ determined from (\ref{Fsigmadamped}) and (\ref{dotfDamped}).
For a given $\alpha(t,s)$, the non-Markovian QSD equation 
(\ref{NMQSDsigmadamped}) can be solved numerically,
having $F(t)$ determined numerically from (\ref{dotfDamped}). 
Note that in the Markov limit, the correlation function $\alpha(t,s)$ 
tends to the Dirac function $\delta(t-s)$. Consequently, $F(t)$ tends 
to the constant $\half f(t,t)=\frac{\lambda}{2}$ and one recovers the 
standard Markov QSD eq. (\ref{MQSD}).

It turns out that non-Markovian QSD can exhibit remarkable properties, 
unknown in the Markov
theory. In order to highlight these features, we proceed analytically 
and assume exponentially decaying environment correlations
$\alpha(t,s)=\frac{\gamma}{2}e^{-\gamma|t-s|-i\Omega(t-s)}$. 
Then we see from (\ref{Fsigmadamped}) and (\ref{dotfDamped}) that
the relevant function $F(t)$ in (\ref{NMQSDsigmadamped}) satisfies
\beq
\dot F(t)=-\gamma F(t) + i(\omega-\Omega)F(t) + \lambda F(t)^2 
+ \frac{\lambda\gamma}{2}
\label{dotF}
\eeq
with initial condition $F(0)=0$. 
With $\tilde\gamma\equiv\gamma-i(\omega-\Omega)$
the solution reads
\beq
F(t)=\frac{\tilde\gamma}{2\lambda}-\frac{\sqrt{\tilde\gamma^2
-2\tilde\gamma\lambda^2}}
{2\lambda} \tanh\left(\frac{t}{2}\sqrt{\tilde\gamma^2
-2\tilde\gamma\lambda^2}
+\mbox{arctanh}\left(\frac{\tilde\gamma}{\sqrt{\tilde\gamma^2
-2\tilde\gamma\lambda^2}}\right)\right).
\eeq

For the remainder of this section we assume exact resonance: 
$\Omega=\omega$ and thus $\tilde\gamma = \gamma$.
Let us first consider the case of short memory or weak coupling, 
$\gamma > 2\lambda^2$. 
For long times, $F(t)$ tends to
$\left(\gamma-\sqrt{\gamma^2-2\gamma\lambda^2}\right)/(2\lambda)$.  
For large
$\gamma$ this asymptotic value tends to $\frac{\lambda}{2}$, which 
corresponds to the Markov limit (\ref{dpsiMarkov}), as it should. 

More interesting, let us consider the opposite case of a long memory
or strong coupling,
$\gamma < 2\lambda^2$. In this case, $F(t)$ diverges to infinity when 
the time $t$
approaches the critical time
$t_c=\left(\pi+2\arctan(\gamma/\sqrt{2\lambda^2\gamma-\gamma^2})\right)/
\sqrt{2\lambda^2\gamma-\gamma^2}$.
What happens is that at time $t_c$, the first component of the vector
$\psi_t$ vanishes,
hence $\sigma_-\psi_{t_c}=0$ and eq.(\ref{dotfDamped}) no longer holds. 
Indeed, the second term of eq.(\ref{NMQSDsigmadamped}) becomes dominant 
and drives the spin to the ground state in a finite time, which we
prove below in terms of the density matrix. In Figure 2a (for individual
trajectories) and Figure 2b (for the ensemble average over 10000 runs)
we see this
effect from solving the non-Markovian QSD equation numerically,
where we choose $\lambda^2 = \Omega = \omega$, so that 
$\omega t_c = \frac{3}{2}\pi \approx 4.71$.
For $t>t_c$ the state $\psi_t$ is constant. This is an example 
where a stationary solution is reached after a finite time! It is the 
first example of a diffusive stochastic Schr\"odinger equation which is 
at the
same time compatible with the no-signaling constraint (ie the evolution of
mixed states depends only on the density matrix, not on a particular 
decomposition into a mixture of pure states) and has no "tails" 
(does not take an infinite time to reach a definite state), see the 
discussions in \cite{PRLcommentsGisinPearle,PearleErice}. 
In \cite{Pearle86Gisin90} it is proven that such a feature is impossible for 
Markov situations. Notice that this peculiar feature holds at 
resonance only. 

Finally, we note that for the intermediate case
$\gamma=2\lambda^2$, one has
$F(t)=\frac{\lambda^3 t}{1+\lambda^2 t} \longrightarrow\lambda$ for
$t\rightarrow \infty$, again approaching a constant value
(the reader may find it helpful to adopt our convenient convention for
the choice of units:
$[z_t]=[\lambda]=[f(t)]=[F(t)]=[1/\sqrt{t}]$ and $[\alpha(t,s)]=[1/t]$).

In order to determine the corresponding master equation for the reduced density
operator, we solve the linear QSD eq. (\ref{LQSD}) where we
make use of the change of variable: $\phi_t\equiv
e^{-i\frac{\omega}{2}\sigma_zt+\lambda\sigma_+\sigma_-\int_0^tF(s)ds}
\psi_t$.
After some computation and taking the ensemble mean analytically, one gets
\beq
\rho_t\equiv M\big[|\psi_t\rangle\langle\psi_t|\big]=
\pmatrix{\rho_{11}(0)e^{-\int_0^t(F(s)+F(s)^*)ds} & 
\rho_{12}(0)e^{-i\omega t-\int_0^t F(s)ds}  \cr \rho_{21}(0)
e^{i\omega t-\int_0^t F(s)^*ds} &
1-\rho_{11}(t)}.
\eeq
This proves that whenever Re$\left(\int_0^tF(s)ds\right)$ diverges for 
a finite time, the density matrix
$\rho_t$ reaches the ground state in that finite time and thus all pure 
state samples have to do so as well.
For the time evolution of this reduced density matrix one gets
\beqa
\dot\rho_t&=&-i\frac{\omega}{2}[\sigma_z,\rho_t] +
\lambda (F(t)+F(t)^*)\left(\sigma_-\rho_t\sigma_+ -
\half\{\sigma_+\sigma_-,\rho_t\}\right) \\
&=&-i\frac{\omega}{2}[\sigma_z,\rho_t] + \int_0^t\K(t,s)\rho_s ds,
\eeqa
where the "memory super-operator" $\K(t,s)$ acts as follows on any 
operator $A$:
\beqa
\K(t,s) A = -\frac{\lambda^2}{2}(\alpha(t,s)+\alpha(t,s)^*)\Big(&&
2e^{-\lambda\int_s^t F(u)du}\sigma_- A\sigma_+ - \{\sigma_+\sigma_-,A\}\\
\nonumber
&-&2(e^{-\lambda\int_s^t F(u)du}-1)\sigma_+\sigma_- 
A \sigma_+\sigma_- \Big)
\eeqa
In Figures 2a and 2b we illustrate this example 
($\lambda^2=\gamma = \Omega = \omega$) for exponentially decaying 
correlations. All individual non-Markovian quantum trajectories reach the
ground state in the critical time $\omega t_c \approx 4.71$ (Fig 2a). 
Taking the ensemble mean over $10000$ trajectories, 
we find very good agreement with the analytical
result of the reduced density matrix (Fig. 2b).

\section{More examples}\label{moreex}

\subsection{Model of energy measurement}
This case, $H=L=L^\dagger$, is a straightforward generalization of 
section \ref{sigma_z}. Again, the environment correlation $\alpha(t,s)$
can be chosen arbitrary. We find $\hat O = H$ in (\ref{Ansatz})
and
the non-Markovian QSD equation (\ref{NMQSD}) for the normalized states 
reads
\beqa\label{NMQSDH}
\frac{d}{dt}\tilde\psi_t &=& -iH\psi_t 
- (H^2-\langle H^2\rangle_t)\tilde\psi_t
\int_0^t\alpha(t,s)ds \\ \nonumber
&+& (H - \langle H\rangle_t)\tilde\psi_t
\left(z_t+\int_0^t\alpha(t,s)^*\langle H\rangle_s ds
+\int_0^t\alpha(t,s)ds\;\;\langle H\rangle_t\right).
\eeqa
For the corresponding master equation we find
\beq
\dot\rho_t=-\int_0^t\alpha(t,s)ds~H[H,\rho_t]-\int_0^t\alpha(t,s)^*ds~
[\rho_t,H]H,
\eeq
hence Tr$(H\rho_t)$ is constant, contrary to the individual expectation 
values $\langle H\rangle_{\tilde\psi_t}$.

The eigenvectors of H are stationary solutions of the non-Markovian
QSD eq.(\ref{NMQSDH}). 
Thus, if the noise is large enough, all initial states tend 
asymptotically to such an eigenstate, as in Markov QSD. However, 
if the noise has long memory, as for example in the extreme case 
of periodic systems (see section \ref{cats}), such a reduction 
property clearly 
does not hold. The exact conditions under which eq. (\ref{NMQSDH})
describes reduction (localization) to eigenstates are not known.
Notice however that if the correlation decays smoothly such that 
$\int_0^t\alpha(t,s)ds$ tends for large times $t$ to a real constant, and 
if $\langle H\rangle_t$ converges for 
large times to a fixed value, then the non-Markovian equation
(\ref{NMQSDH}) tends to
\beq\label{NMQSDHappr}
\frac{d}{dt}\tilde\psi_t = -iH\tilde\psi_t - (H^2-\langle H^2\rangle_t)
\tilde\psi_t const 
+(H - \langle H\rangle_t)\tilde\psi_t
\left(z_t+const\langle H\rangle_t\right).
\eeq
The long time solutions of this equation are the same as the long time 
solutions of the corresponding Markov approximation. The latter is 
the Markov QSD equation, 
hence the asymptotic solutions tend to eigenstates of $H$.
The previous subsection \ref{sigma_z} provides an example of this more 
general statement for $H=\frac{\omega}{2}\sigma_z$.

\subsection{A simple toy model}
In this subsection we use a simple toy model \cite{Hepp} to illustrate 
that the non-Markovian QSD
equation (\ref{NMQSD}) may contain unexpected additional terms that 
cancel in the Markov limit.
Consider $H=p$ and $L=q$ and an arbitrary environment correlation function
$\alpha(t,s)$. Then the Ansatz (\ref{Ansatz}) for replacing 
the functional
derivative with some operator satisfying the consistency 
condition (\ref{consistentAnsatz}) reads
\beq
\frac{\delta\psi_t}{\delta z_s} = (q-(t-s))\psi_t.
\eeq
Thus, the non-Markovian QSD eq. (\ref{NMQSD}) takes the form
\beqa\label{NMQSDhepp}
\frac{d}{dt}\tilde\psi_t&=& -ip\tilde\psi_t 
- (q^2-\langle q^2\rangle_t)\int_0^t\alpha(t,s)ds\tilde\psi_t \\ \nonumber
&+&(q - \langle q\rangle_t)\tilde\psi_t
\left(z_t+\int_0^t\alpha(t,s)^*\langle q\rangle_s  ds
+\int_0^t\alpha(t,s)ds\langle q\rangle_t\right) \\ \nonumber
&+&(q - \langle q\rangle_t)\tilde\psi_t \int_0^t(t-s)\alpha(t,s)ds.
\eeqa
The first two lines of this non-Markovian QSD equation could have been
expected, since they have the same form as in the previous examples, see 
for instance eq. (\ref{NMQSDH}). The last line of the above equation, however, 
has no counterpart in the previous examples. Clearly, it vanishes in the 
Markov limit $(\alpha(t,s)\rightarrow \delta(t-s))$,
when the non-Markovian QSD equation (\ref{NMQSDhepp}) for this model
reduces to the Markov QSD equation (\ref{MQSD}).

\subsection{Quantum Brownian motion model}
In this subsection we consider the important case of quantum Brownian 
motion of a harmonic oscillator \cite{qbm}, that is we choose 
$H = \frac{\omega}{2}(p^2 + q^2)$, $L=\lambda q$, and arbitrary 
environmental correlation $\alpha(t,s)$. As shown in Appendix C, the
basic linear non-Markovian QSD equation for this quantum Brownian motion case
is again the fundamental linear equation (\ref{LQSD}).

It turns out that the functional derivative in (\ref{LQSD}) is more 
complicated in this case, because $\hat O(t,s,z)$ depends explicitly on the 
noise $z$. However,
fortunately, this dependence is relatively simple. Indeed, let
\beq\label{qbmansatz}
\frac{\delta\psi_t}{\delta z_s} \equiv \hat O(t,s,z)\psi_t =
\left[f(t,s)q + g(t,s) p + i\int_0^t ds' j(t,s,s') z_{s'}\right] \psi_t.
\eeq
The consistency condition
(\ref{consistentAnsatz}) leads to the following equations for the
unknown functions $f(t,s),g(t,s)$ and $j(t,s,s')$ in (\ref{qbmansatz}):
\beqa\label{qbmeqs}
\partial_t f(t,s) & = & \omega g(t,s) + i f(t,s)
\int_0^t ds' [\alpha(t,s')g(t,s')]\\ \nonumber
& & -2ig(t,s)\int_0^t ds' [\alpha(t,s')f(t,s')] 
-i\int_0^t ds' [\alpha(t,s')j(t,s',s)]\\
\partial_t g(t,s) & = & -\omega f(t,s) - i g(t,s)
\int_0^t ds' [\alpha(t,s')g(t,s')]\\
j(t,s,t) & = & g(t,s) \\
\partial_t j(t,s,s') & = & -i g(t,s)
\int_0^t ds'' [\alpha(t,s'')j(t,s'',s')].
\eeqa
These equations have to be solved together with the non-Markovian
QSD equation (\ref{NMQSD}).

If, for simplicity, we assume exponentially decaying environment 
correlations $\alpha(t,s)=\frac{\gamma}{2}e^{-\gamma|t-s|}$
and introducing capital letters for the integrals, 
$X(t)\equiv\int_0^t\alpha(t,s)x(t,s)ds$, 
for $x=f,g,j$, one obtains the simpler closed set of equations
\beqa
\dot F(t)&=&\frac{\lambda\gamma}{2}-\gamma F(t) + \omega G(t) 
- i\lambda F(t)G(t) -i\lambda\tilde J(t) \\
\dot G(t)&=&-\gamma G(t)-\omega F(t)-i\lambda G(t)^2 \\
\dot{\tilde J}(t)&=&\frac{\lambda\gamma}{2}G(t) - 2\gamma\tilde J(t) 
- i\lambda G(t)\tilde J(t),
\eeqa
where $\tilde J(t)\equiv\int_0^t\alpha(t,s')J(t,s')ds'$. 
The initial conditions read
$F(0)=G(0)=\tilde J(0)=0$. Finally, $J(t,s)$ can be determined from 
the solutions of the above equations, we get
\beq
J(t,s)=\lambda G(s)e^{-\int_s^t(\gamma+i\lambda G(s'))ds'}.
\eeq
Hence, the non-Markovian QSD equation for quantum Brownian motion becomes
\beqa \label{NMQSDQBM}
\frac{d}{dt}\tilde\psi_t&=& -iH\tilde\psi_t - (q^2-\langle q^2\rangle_t)
\tilde\psi_t F(t) 
- \big(qp-\langle qp\rangle_t-p\langle q\rangle_t+ \langle p\rangle_t
\langle q\rangle_t \big)\tilde\psi_t G(t) \\ \nonumber
&+&(q - \langle q\rangle_t)\tilde\psi_t
\left(z_t+\int_0^t\alpha(t,s)^*\langle q\rangle_s  ds 
+ \langle q\rangle_t F(t)
-i\int_0^t J(t,s')\big(z_{s'}+\int_0^{s'}\alpha(s',s)^*\langle q\rangle_s  
ds\big)ds' \right).
\eeqa
Let us make some comments about this non-Markovian QSD equation. First, recall 
that it corresponds to the exact solution of the quantum Brownian motion 
problem 
\cite{qbm} of a harmonic oscillator. 
Next, this example shows a new feature that we didn't encounter in the 
previous examples: the noise $z_t$
enters the equation non-locally in time. Thirdly, terms involving the
operator $qp$ appear, although
there are no such terms either in the Hamiltonian or in the environment 
operator $L=\lambda q$.
Finally, since this equation is exact, it is a good starting point to 
tackle the quantum Brownian motion problem using this new approach
and to find its proper Markov limit. 
In connection with this last point, we
emphasize that the master equation corresponding to eq. (\ref{NMQSDQBM}) 
necessarily preserves
positivity \cite{Ambega91}  because it provides a decomposition of the 
density operator into pure states at all times.
However, these questions and numerical simulations are left for future 
work.

\subsection{Harmonic oscillator at finite temperature}

As another important example of an open quantum system we briefly sketch 
the case of
a harmonic oscillator $H=\omega a^\dagger a$ coupled to a finite
temperature environment through $L_-=\lambda_- a$. As explained in detail 
in Appendix C, the finite temperature also induces absorption from the
bath, which has to be described by a second environment operator 
$L_+=\lambda_+ a^\dagger$.
Hence, the linear non-Markovian QSD equation (\ref{LQSD}) has to be modified
and involves two independent noises $z^-_t$ and $z^+_t$,
\beq\label{LQSDa}
\frac{d}{d t}\psi_t=-iH\psi_t 
+ \lambda_-a\psi_t z^-_t - 
\lambda_-a^\dagger\int_0^t \alpha^-(t,s)\frac{\delta\psi_t}{\delta z^-_s}ds
+ \lambda_+a^\dagger\psi_t z^+_t - 
\lambda_+a \int_0^t \alpha^+(t,s)\frac{\delta\psi_t}{\delta z^+_s}ds,
\eeq 
see eq. (\ref{LQSDTGT0}) in Appendix C.
This equation can be solved with the following Ans\"atze:
\beq
\frac{\delta\psi_t}{\delta z^-_s} =
\left[f_-(t,s)a + \int_0^t ds' j_-(t,s,s') z^+_{s'}\right] \psi_t
\eeq
\beq
\frac{\delta\psi_t}{\delta z^+_s} =
\left[f_+(t,s)a^\dagger + \int_0^t ds' j_+(t,s,s') z^-_{s'}\right] \psi_t.
\eeq
Using similar techniques as in the previous subsection, the evolution 
equations for $f_\pm(t,s)$ and $j_\pm(t,s,s')$ can be obtained and thus 
the resulting non-Markovian
QSD equation can be written in closed form. A new feature of this example, 
again unknown in the Markov case, is that each of the two environment 
operators $L_-$ and $L_+$, is coupled to both noises.

\section{Harmonic oscillator coupled to a few oscillators: decay and
revival of Schr\"odinger cat states}\label{cats}
The case of a harmonic oscillator coupled to a finite or infinite number of
harmonic oscillators all of which are initially in their ground state (zero temperature),
$H=\omega a^\dagger a$, $L = \lambda a$, is very similar to the damped
spin $\half$ example treated in subsection \ref{DampedSpin}. The Ansatz 
$\frac{\delta\psi_t}{\delta z_s}=f(t,s)a\psi_t$ similar to 
(\ref{AnsatzDamped}) 
holds with $f(t,s)$ and $F(t)$ satisfying the same equations 
(\ref{dotfDamped}) and (\ref{Fsigmadamped}).
Thus, the non-Markovian QSD equation (\ref{NMQSD}) for this situation reads
\beq\label{NMQSDzerot}
\frac{d}{dt}\tilde\psi_t 
= -i\omega a^\dagger a\tilde\psi_t + (a-\langle a \rangle_t)\tilde\psi_t
(z_t+\int_0^t\alpha^*(t,s) \langle a^\dagger\rangle_s ds
                           +\lambda F(t)\langle a^\dagger\rangle_t)
 - \lambda F(t) (a^\dagger a - \langle a^\dagger a\rangle_t)\tilde\psi_t .
\eeq

Again, this non-Markovian QSD equation reduces to the 
Markov equation (\ref{MQSD}) for $\alpha(t,s) = \delta(t-s)$ since
in this case $F(t)=\frac{\lambda}{2}$ according to (\ref{Fsigmadamped}).
As in the case of a dissipative spin (subsection \ref{DampedSpin}), 
for exponentially
decaying bath correlations at resonance, the system oscillator
may reach its ground state in a finite time,
provided the correlation time $1/\gamma$ is long enough.

Notice also that (\ref{NMQSDzerot}) preserves coherent states 
$|\beta\rangle$. The time evolution of the complex number $\beta_t$ labeling 
these coherent states is given by
\beq
\dot\beta_t=\big(-i\omega-F(t)\big)\beta_t.
\eeq

More interesting than a coherent state initial condition is the case of a 
superposition $|\beta\rangle + |-\beta\rangle$
of two symmetric coherent states, known as a 'Schr\"odinger cat' 
\cite{cat}. If the correlation
decays, so does the 'cat'. If, in contrast,
the environment consists only of a finite number of
oscillators, then the 'cat' will first decay, due to the localization 
property of QSD,
but since the entire system is quasi-periodic, the 'cat' will then revive! 

As an illustration, we simulate the extreme case where the 'environment'
consists of only a single oscillator. It thus models the decay and revival
of a field cat state in a cavity that is isolated from the outside, but 
coupled 
to a second cavity, to which it may decay reversibly. Such
an experiment on reversible decoherence was proposed recently
in \cite{Haroche}.
In this simple case, the environment correlation function reads
\beq
\alpha(t,s)=e^{-i\Omega(t-s)},
\eeq
where $\Omega$ is the
frequency of the single `environment' oscillator.
Figure 3 shows the time evolution of the $Q$-function of such a 
'Schr\"odinger cat' in phase space for $\Omega = 0.5\omega$ and
a coupling strength between the two oscillators of $0.1\omega$.
Apart from an overall oscillatory
motion due to the `system' Hamiltonian $\omega a^\dagger a$, we see 
how the cat
first decays but later becomes alive again. Further investigations of 
stochastic state vector descriptions of such
reversible decoherence processes are left for future investigations.
It is worth mentioning that depending on the stochastic process,
the cat my subsequently decay into either of its two components.

\section{Shifting the system-environment boundary}
In this section we consider a situation where the 'Heisenberg cut' between 
the system and the environment is not obvious. Since the non-Markovian 
QSD equation provides the exact solution of the total system-environment 
dynamics, the description of the system does not depend on this cut.
This is in contrast to the usual Markov approximation, where the position 
of the cut is crucial. As an example,
let us consider a system consisting of one spin $\half$ and one harmonic 
oscillator,
the two subsystems being linearly coupled. Assume moreover that the spin 
$\half$ is coupled
to a heat bath at zero temperature, see figure 4. The total Hamiltonian 
reads:
\beq
H_{total}=H_1 + H_2 + H_{12} + H_{env} + H_I
\eeq
with
\beqa
H_1&=&\frac{\omega_1}{2} \sigma_z \\
H_2&=&\omega_2 a^\dagger a \\
H_{12}&=&\chi(\sigma_-a^\dagger+\sigma_+ a) \\
H_{env}&=&\sum_\omega \omega a_\omega^\dagger a_\omega \\
H_I&=&\sum_\omega \chi_\omega(\sigma_-a_\omega^\dagger+\sigma_+ a_\omega).
\eeqa

We can either consider the spin-oscillator system coupled to a heat bath, 
or consider only
the spin coupled to a heat bath and coupled to an auxiliary oscillator, as
illustrated in Figure 4. In the 
first case, we can consider the Markov QSD description, ie a family of 
spin-oscillator
state vectors $\psi_t(\xi)$ indexed by the complex Wiener processes
$\xi_t$.
In the second case, using non-Markovian QSD we have a family of spin 
$\half$ state vectors
$\phi_t(\xi,z)$ indexed by the same $\xi_t$ plus the non-Markovian 
noise $z_t$ with correlations
\beq
M[z_t^*z_s] = e^{-i\omega_2 (t-s)},
\eeq
The (linear) stochastic eqs.(\ref{LQSD}) governing $\psi_t$ and $\phi_t$ 
read
\beqa
\dot\psi_t&=&-i(H_1+H_2+H_{12})\psi_t + \lambda \sigma_-\psi_t\xi_t 
- \frac{\lambda^2}{2}\sigma_+\sigma_-\psi_t \\
\dot\phi_t&=&-iH_1\phi_t + \lambda \sigma_-\phi_t\xi_t 
- \frac{\lambda^2}{2}\sigma_+\sigma_-\phi_t
+ \chi\sigma_-\phi_t z_t - \chi\sigma_+\int_0^t e^{-i\omega_2 (t-s)}
\frac{\delta\phi_t}{\delta z_s}ds
\label{dphi}
\eeqa
where $\lambda$ is a function of the $\chi_\omega$'s, 
that is of the strength of the spin-heat-bath coupling.

A natural question in the present framework is to study the 
"Heisenberg cut": compare the states of the spin $\half$ averaged over the 
noise $z$ with the mixed state obtained by tracing out the second oscillator 
(Tr$_2$) from the 1-oscillator-spin states, i.e. we ask whether the 
equality
\beq
M_z\left[|\phi_t(\xi,z)\rangle\langle\phi_t(\xi,z)|\right]
\buildrel ?\over {=} 
Tr_2(|\psi_t(\xi)\rangle\langle\psi_t(\xi)|)
\label{equ}
\eeq
holds.
According to the general non-Markovian QSD theory presented in this paper,
the spin$\half$ state should be independent of the position
of the Heisenberg cut. Below we illustrate this feature using the 
present example.

By assumption the oscillator starts in the ground state: 
$\psi_0=\phi_0\otimes|0\rangle$. Hence, the state $\psi_t$ can be expanded 
as
\beq
\psi_t = c_0(t)|\da,0\rangle + c_1(t)|\ua,0\rangle + c_2(t)|\da,1\rangle,
\eeq
where
\beqa
\dot c_0&=&\lambda\xi(t) c_1 + i\frac{\omega_1}{2}c_0 \label{dc0} \\
\dot c_1&=&-(i\frac{\omega_1}{2}+\frac{\lambda^2}{2})c_1 - i\chi c_2 
\label{dc1} \\
\dot c_2&=&-i((\omega_2-\frac{\omega_1}{2})c_2 + \chi c_1) \label{dc2}.
\eeqa
Tracing out the single harmonic oscillator, one obtains the spin $\half$ state 
(in the $\ua\da$ basis)
\beq
\rho_1\equiv Tr_2(|\psi_t(\xi)\rangle\langle\psi_t(\xi)|)=\pmatrix{|c_1|^2 
& c_0^*c_1 \cr c_0c_1^* & |c_0|^2+|c_2|^2}.
\eeq

We now turn to the alternative description of the same situation, but 
with the 'cut' 
between the spin $\half$ and the oscillator. In order to solve eq. 
(\ref{dphi}) we make the usual Ansatz
\beq
\frac{\delta\phi_t}{\delta z_s}=f(t,s)\sigma_-\phi_t,
\eeq
where the consistency condition (\ref{consistentAnsatz}) leads to 
$\partial_t f(t,s)=(i\omega_1+\frac{\lambda^2}{2}+\chi F(t))f(t,s)$,
where $f(t,t)=\chi$ and $F(t)=\int_0^t\alpha(t,s)f(t,s)ds$. Consequently,
\beq
\dot F(t)=\chi + (i\omega_1-i\omega_2+\frac{\lambda^2}{2}+\chi F(t))F(t).
\label{dF}
\eeq
Using the notations $\phi_t=v_0(t)|\da\rangle+v_1(t)|\ua\rangle$ one gets
\beqa
\dot v_0&=&i\frac{\omega_1}{2}v_0 + (\lambda\xi_t+\chi z_t)v_1 
\label{dv0} \\
\dot v_1&=&-(i\frac{\omega_1}{2}+\frac{\lambda^2}{2}+\chi F(t))v_1. 
\label{dv1}
\eeqa
Note that since $\dot v_1$ is independent of $z_t$, $v_1(t)$ is itself 
independent of $z$, hence,
\beq
\frac{d}{dt}M_z[v_0]=i\frac{\omega_1}{2}M_z[v_0] + \lambda\xi_t v_1 .
\eeq
Averaging over the z-noise, one obtains the spin $\half$ state 
(in the  $\ua\da$ basis)
\beq
\rho_2\equiv M_z\left[|\phi_t(\xi,z)\rangle\langle\phi_t(\xi,z)|\right]
=\pmatrix{|v_1|^2 & M_z[v_0^*]v_1 \cr M_z[v_0]v_1^* & M_z[|v_0|^2]}.
\eeq
Finally, a straightforward comparison of eqs. 
(\ref{dc0},\ref{dc1},\ref{dc2}) and (\ref{dF}, \ref{dv0},\ref{dv1}) 
shows that $c_0=M_z[v_0]$, $c_1=v_1$ and $c_2=-iFv_1$.
Hence, 3 of the 4 entries of the matrices $\rho_1$ and $\rho_2$ are equal. 
The equality of the fourth entry follows from the general feature that 
linear non-Markovian QSD preserves the mean of the square norm.

This completes the proof that $\rho_1=\rho_2$: the spin $\half$ state 
is independent of 
the position of the Heisenberg cut, for all times and all realizations 
of the heat bath induced noise $\xi$.
This illustrates the general fact that non-Markovian QSD attributes 
stochastic pure states to
systems in a way which depends on the position of the Heisenberg cut, 
but which is
consistent for all possible choices of the cut. See Fig 4. for the
illustration of these relationships. This is in opposition to the case
prevailing in Markovian unravellings.

\section{Open problems}\label{prob}
This paper is the first presentation of non-Markovian QSD. Admittedly, there 
remain many open questions and a lot of work has still to be done to 
exploit all the possibilities opened up by this new approach. In this section 
we list some of the open problems:
\begin{enumerate}
\item
The ultimate goal would be to develop a general purpose numerical 
simulation program.  However, at present no general recipe is known. 
\item
When do the long time limit and the Markov limit commute? A question which 
is of particular interest for quantum Brownian motion.
\item
If the initial condition is not factorized, the present approach must 
be generalized.
\item
In the Markov case unravellings exist both with continuous trajectories 
and with quantum jumps. In the non-Markovian case, the only unravelling 
known at present is the continuous non-Markovian QSD described in this 
article. What about non-Markovian unravellings with quantum jumps?
\item
In the Markov case, continuous QSD unravellings exist for real or pure
imaginary noise, as well as for complex noise. What about the 
non-Markovian case? It seems that in the present case complex noise is 
essential.
\item
Note that most of the non-Markovian master equations used in this article 
have known analytical solutions. In these cases, 
the general Zwanzig form \cite{Zwanzig} of the master equation:
\beq
\dot\rho_t=\int_0^t\K(t-s)\rho_s ds
\eeq
with the memory kernel $\K(t-s)$ could be rewritten as a Lindblad type 
master equation with time-dependent coefficients. Then, the master 
equation can also be simulated using Markov QSD with time dependent 
coefficients. However, if the solution
of the master equation is not known explicitly,
or does not lead to a Lindblad type equation,
then numerical simulation has to use the non-Markovian QSD theory. 
It would be interesting to illustrate non-Markovian QSD for more of 
such examples and to study the conditions under which a non-Markovian
problem can be treated with Markovian unravellings.
\item
How does non-Markovian QSD compare with consistent histories
\cite{Diosi95} and other approaches?
For instance, it was shown in \cite{Diosi96b} that the 
solutions of the non-Markovian
eq. (\ref{NMQSD}) can be considered as conditional states in the 
framework of a "hybrid"
representation of the fully quantized microscopic system.
\item
What is the perturbation expansion of the non-Markovian QSD eq. 
(\ref{NMQSD}) in terms of the memory time $\gamma^{-1}$? The zeroth 
order term would be the Markov
QSD eq. (\ref{MQSD}), what about the higher orders?
\item
Finally, non-Markovian QSD should be applied to open problems in physics, 
where non-Markovian effects are relevant,
such as semi-conductor lasers \cite{Imamog94}, or atom lasers 
\cite{Savage}.
\end{enumerate}

\section{Conclusion}\label{concl}
We present a stochastic equation for pure states describing non-Markovian 
quantum state diffusion,
compatible with non-Markovian master equations. We illustrate its 
power with several examples. In essence, we show that quantum
(finite or infinite) harmonic oscillator environments 
can be modeled by classical, complex 
Gaussian processes, entering the non-linear, non-Markovian stochastic
Schr\"odinger equation for the `system' state which we derive in this 
paper.

Several authors have proposed stochastic pure-state descriptions of
such non-Markovian situations 
using fictious modes added to the system in such a way 
as to make to dynamics of the enlarged hypothetical system Markovian 
\cite{Imamog94,Garraw97}. Others, \cite{Bay97} treat a non-Markovian
problem with an explicitly time-dependent Markov unravelling.
In our 
approach, by contrast, there are no additional modes, hence the
system is as small as possible, and the stochastic
Schr\"odinger equation becomes genuinely non-Markovian.
This is of interest for efficient 
numerical simulation and high-focus insight into the relevant physical 
processes.
Also, non-Markovian quantum trajectories are in general much smoother  
than those of Markov processes, which might even help to reduce further
the numerical effort.

Let us stress an important conceptual difference between Markov QSD
and non-Markovian QSD. In the Markov case, one starts from a master 
equation for mixed states and associates to it a stochastic Schr\"odinger 
equation. The master equation may either be derived from a microscopic 
model, or merely be based on phenomenological motivations \cite{QSD3}. 
In the non-Markovian case, on the contrary, one starts from the 
stochastic Schr\"odinger eq. (\ref{LQSD}). The existence of a master 
equation is guaranteed by the microscopic model summarized in Appendix C.
In general, however, the explicit form of this master equation is not known. 
Nevertheless, this existence ensures that the corresponding stochastic 
Schr\"odinger equation 
for normalized states (\ref{NMQSD}) does not allow 
arbitrary fast signaling, despite its nonlinearity \cite{GisRig95}. 

From a pragmatic point of view, the Hamiltonian and environment operators 
in eq.(\ref{LQSD}) can either be derived from a microscopic
theory, or be merely based on phenomenological motivations. 
Non-Markovian master equations are almost always exceedingly difficult 
to treat, even numerically. However, one can always start from the 
non-Markovian QSD approach of this paper, which appears thus more 
fundamental than 
the master equation approach.

\section*{Acknowledgments}
We thank IC Percival for helpful comments and the University of Geneva 
where part of the work was done.
LD is supported by the Hungarian Scientific Research Fund through grant
T016047.
NG thanks the Swiss National Science Foundation.
WTS would like to thank
the Deutsche Forschungsgemeinschaft for support through the SFB 237
"Unordnung und gro{\ss}e Fluktuationen".

\begin{appendix}

\section{Frequency representation}
It is sometimes useful to express the noise by frequency components 
$z_\omega$:
\beq\label{zomega}
z_t=\sum_\omega z_\omega e^{i\omega t},
\eeq
where the frequencies $\omega$ can take positive as well as negative 
values.
Also the correlation function can be written in Fourier representation:
\beq\label{alphaomega}
\alpha(t,s)=\alpha(t-s) = 
\sum_\omega \alpha_\omega e^{-i\omega(t-s)},~~~~~~\alpha_\omega > 0. 
\eeq
The correlation of the Fourier components of the noise is trivial:
$
M[z_\omega^\star z_\lambda]=\delta_{\omega\lambda}\alpha_\omega.
$
In this representation the distribution functional becomes a simple 
Gaussian distribution over all $z_\omega$'s:
\beq\label{Pzfreq}
P(z)={\cal N}\exp
\left( -\sum_\omega\frac{\vert z_\omega \vert^2}{\alpha_\omega} \right)
\eeq
and the states $\psi_t$ become functions of the frequency amplitudes
$z_\omega$ of the noise.
We can then write the fundamental linear non-Markovian QSD eq. (\ref{LQSD}) 
in terms of them:
\beq\label{NMQSDfreq}
\frac{d}{d t}\psi_t=-iH\psi_t + \sum_\omega 
\Bigl( Le^{i\omega t}z_\omega 
- L^\dagger\alpha_\omega e^{-i\omega t}\frac{\partial}{\partial z_\omega} 
\Bigr)\psi_t.
\eeq
This frequency representation is a helpful tool to discuss the mathematical 
properties of the non-Markovian stochastic Schr\"odinger equation 
(\ref{NMQSD}), as we do in Appendices B and C.
Remember that in eq. (\ref{NMQSDfreq}) we assume the 
initial condition to be independent of the noise: $\psi_0(z)=\psi_0$.

\section{Girsanov transformation for non-Markovian QSD}
As time goes by, Girsanov transformation distorts the distribution $P(z)$ 
(\ref{Pzfreq}) of the complex noise into $\tilde P_t(z)$ according to
eq.(\ref{rhonormpsi}).  In frequency representation, we have
\beq
\tilde P_t(z)={\cal N}\Vert \psi_t(z)\Vert^2 \exp
\left( -\sum_\omega \frac{\vert z_\omega \vert^2}{\alpha_\omega} \right).
\label{Ptz}
\eeq
We assume that at $t=0$ the state $\psi_0$ is normalized and does not 
depend on $z$. So, initially, $\tilde P_0(z)$ is identical with $P(z)$.

We find the time evolution of $\tilde P_t(z)$ from the linear 
non-Markovian Schr\"odinger equation (\ref{LQSD}) in frequency 
representation (\ref{NMQSDfreq}). Using eq.(\ref{Ptz}), we find
\beq
\frac{d}{dt}\tilde P_t(z) 
={\cal N}\langle\psi_t(z)\vert\frac{d}{dt}\psi_t(z)\rangle
\exp\left( -\sum_\omega \frac{\vert z_\omega \vert^2}{\alpha_\omega} 
\right) + c.c.
\label{dotPtz}
\eeq
Now we make a crucial observation.
The solution $\psi_t(z)$ of (\ref{NMQSDfreq}), with initial condition 
$\psi_t(z)=\psi_0$, is analytic in all $z_\omega$'s. Then it follows that
$\partial\vert\psi_t(z)\rangle/\partial z_\omega^\star=
\partial\langle\psi_t(z)\vert/\partial z_\omega=0$ for all $z_\omega$. 
Hence, when inserting eq. (\ref{NMQSDfreq}) into eq. (\ref{dotPtz}), 
we can substitute
\beq
\langle\psi_t(z)\vert 
L^\dagger\frac{\partial}{\partial z_\omega}\psi_t(z)\rangle
= \frac{\partial}{\partial z_\omega}
\langle L^\dagger\rangle_t \Vert\psi_t(z)\Vert^2, 
\label{analtrick}
\eeq
and we obtain 
\beq
\frac{d}{dt}\tilde P_t(z)=
-\sum_\omega\alpha_\omega e^{-i\omega t} \frac{\partial}{\partial z_\omega} 
 \langle L^\dagger \rangle_t \tilde P_t(z)
                            + c.c.
\label{dotPtzresult}
\eeq
This is a remarkable result. It shows that the Girsanov 
transformation is equivalent to a drift of the random variable $z$.
We read off the drift velocities directly from eq. (\ref{dotPtzresult}):
\beq
\frac{d}{dt}z_\omega=\alpha_\omega e^{-i\omega t}\langle L^\dagger 
\rangle_t.
\label{drift}
\eeq

One can see that the Girsanov transformation preserves the normalization
of the distribution $\tilde P_t(z)$. This has the immediate consequence 
that the non-Markovian stochastic Schr\"odinger equation (\ref{LQSD}) 
preserves the mean norm of the quantum state:
\beq\label{normproof} 
M\left[\Vert\psi_t\Vert^2\right]\equiv\int\Vert\psi_t\Vert^2 P(z)dz=
                                    \int\tilde P_t(z)dz=1.
\eeq
Now we are going to derive the stochastic non-Markovian Schr\"odinger
equation for the normalized states 
$\tilde\psi_t(z)=\psi_t(z)/\Vert\psi_t(z)\Vert$,
where $\psi_t(z)$ is the unnormalized solution of the linear
stochastic equation (\ref{LQSD}). First, we solve the drift eq.
(\ref{drift}) for the trajectories $z_\omega(t)$, with the initial 
conditions $z_\omega(0)=z_\omega$ for all $\omega$:
\beq\label{shiftfreq}
\tilde z_\omega(t) = 
z_\omega + \int_0^t \alpha_\omega e^{-i\omega s}\langle L^\dagger\rangle_s 
ds.
\eeq
where 
$\langle L^\dagger\rangle_t = 
\langle\psi_t(\tilde z(t))|L^\dagger|\psi_t(\tilde z(t))\rangle/
\langle\psi_t(\tilde z(t))|\psi_t(\tilde z(t))\rangle$.
The Girsanov-transformation (\ref{Girs}) leaves invariant the probability
of the noise $z$ along the above trajectories:
\beq\label{invPtilde}
\tilde P_t(\tilde z(t))d\tilde z(t)\equiv P(z)dz 
\eeq
for all $z_\omega$. Hence, we can write the stochastic
unravelling (\ref{rhonormpsi}) as follows:
\beq
\rho_t=\tilde M_t\left[\vert\tilde\psi_t(z)\rangle\langle\tilde\psi_t(z)
\vert\right]
=M\left[\vert\tilde\psi_t(\tilde z(t))\rangle
      \langle\tilde\psi_t(\tilde z(t))\vert\right].
\eeq
The mean value on the very right refers to the simple undistorted 
distribution
$P(z)$. To calculate it, one has to express $\psi_t(\tilde z(t))$ as a 
function
of the initial amplitudes $z_\omega=\tilde z_\omega(0)$. Remember that 
$\psi_t(z)$ is the solution of the linear non-Markovian equation 
(\ref{LQSD}) or (\ref{NMQSDfreq})
with initial condition $\psi_t(z)=\psi_0$. The additional time dependence of
$\psi_t(\tilde z(t))$ through $\tilde z(t)$ appends a new term to the 
evolution equation of these 'Girsanov-shifted' states,
so that we find the following stochastic evolution equation:
\beqa\label{LQSDGIR}
\frac{d}{dt}\psi_t(\tilde z(t))& = &\frac{\partial}{\partial t}\psi_t +
\sum_\omega \dot z_\omega \frac{\partial}{\partial z_\omega} \psi_t\\
& = & -iH\psi_t + \sum_\omega Le^{i\omega t}\tilde z_\omega - 
(L^\dagger-\langle L^\dagger\rangle_t) \int_0^t \alpha(t,s) 
\hat O(t,s,\tilde z)ds\psi_t
\eeqa
where we used (\ref{Ansatz}), (\ref{sLQSD}), and (\ref{drift}).
Finally, these states have to be normalized. The resulting evolution
equation for the normalized states $\tilde\psi_t$ 
is our central result, given by eq.(\ref{NMQSD}). In the time domain,
the shifted noise (\ref{shiftfreq}) takes the form (\ref{shift}).  

\section{Review of the linear non-Markovian theory}
Here we briefly review the microscopic origin of the linear
non-Markovian stochastic Schr\"odinger equation (\ref{LQSD})
- see \cite{Diosi96,Strunz96b,DioStr97}.
The linear non-Markovian QSD equation results from a standard model
of a system interacting with an environment of harmonic oscillators,
represented by a set of bosonic annihilation and creation operators 
$a_\omega, a_\omega^\dagger$.
The interaction term $H_I$ between system and environment is chosen
to be linear in the $a_\omega$s and arbitrary in the system operator $L$:
$H_I = \sum_\omega\chi_\omega (L a_\omega^\dagger + L^\dagger a_\omega)$, 
with some coupling constants $\chi_\omega$. Thus, the model is defined
by
\beqa\label{totalmodel}
H_{tot} & = & H_{sys} + H_I + H_{env} \\
 & = & H_{sys} 
+ \sum_\omega\chi_\omega (L a_\omega^\dagger + L^\dagger a_\omega)
+ \sum_\omega \omega a_\omega^\dagger a_\omega
\eeqa
Solving this total closed system in a clever way leads to the linear 
non-Markovian stochastic Schr\"odinger equation (\ref{LQSD}) for the 
system state
$\psi_t(z)$. As initial condition we assume a factorized form
$\rho_{tot} = |\psi_0\rangle\langle\psi_0|\otimes \rho_T$ 
for the total density operator,
with all bath oscillators initially in some thermal state 
$\rho_T =\otimes_\omega \rho_\omega(T)$.

\subsection{Zero temperature}
In \cite{DioStr97} it was shown that if all the environment oscillators 
are initially
in their ground state $(T=0)$, the dynamics of the reduced density 
operator $\rho_t = \mbox{tr}_{env}\rho_{tot}(t)$ of the model 
(\ref{totalmodel}) 
can be unraveled $(\rho_t = M[|\psi_t(z)\rangle\langle\psi_t(z)|])$ using 
the linear stochastic Schr\"odinger equation (\ref{LQSD}),
\beq\label{LQSDT0}
\frac{d}{d t}\psi_t=-iH\psi_t + L\psi_t z_t - 
L^\dagger\int_0^t \alpha(t,s)\frac{\delta\psi_t}{\delta z_s}ds
\eeq
where the colored complex stochastic processes $z_t$ with zero mean 
satisfy
\beq\label{micro}
M[z_t^* z_s] = \sum_\omega \chi_\omega^2
\;e^{-i\omega(t-s)} \equiv \alpha(t,s)
,\;\;\;M[z_t z_s] = 0.
\eeq
We see the microscopic origin of the bath correlation function
$\alpha(t,s)$ at zero temperature. For real physical systems we have
$\omega>0$ in (\ref{micro}). To model an arbitrary time-translation
invariant correlation function, one needs environment oscillators
with negative frequencies as well.

\subsection{Finite temperature}
In order to derive the linear non-Markovian QSD equation at finite
temperatures, we use a simple mathematical trick, well 
known in field theory \cite{SemUme83}: the non-zero temperature density 
operator $\rho_T$ of the heat bath can be canonically mapped 
onto the zero-temperature 
density operator (the vacuum) of a larger (hypothetical) environment.
The problem at $T>0$ is thus reduced to the problem at $T=0$, whose
linear non-Markovian QSD equation (\ref{LQSDT0}) we already know.
The resulting finite temperature linear non-Markovian QSD equation is
\beq\label{LQSDTGT0}
\frac{d}{d t}\psi_t=-iH\psi_t 
+ L\psi_t z^-_t - 
L^\dagger\int_0^t \alpha^-(t,s)\frac{\delta\psi_t}{\delta z^-_s}ds
+ L^\dagger\psi_t z^+_t - 
L \int_0^t \alpha^+(t,s)\frac{\delta\psi_t}{\delta z^+_s}ds.
\eeq
It thus depends on two independent processes $z^-_t, z^+_t$ 
with zero means and with temperature dependent 
correlations
\beq\label{zetminus}
M[{z^-_t}^* z^-_s] = 
\sum_\omega (\bar n_\omega + 1) \chi_\omega^2
\;e^{-i\omega(t-s)} \equiv \alpha^-(t,s),\;\;\; M[z^-_t z^-_s] = 0
\eeq
and
\beq\label{zetplus}
M[{z^+_t}^* z^+_s] = 
\sum_\omega \bar n_\omega \chi_\omega^2
\;e^{i\omega(t-s)} \equiv \alpha^+(t,s),\;\;\; M[z^+_t z^+_s] = 0.
\eeq
Here, $\bar n_\omega = (\exp\frac{\hbar\omega}{kT}-1)^{-1}$ denotes
the average thermal number of quanta in the mode $\omega$.
We identify these terms as describing the stimulated ($\bar n$) and 
spontaneous ($+1$) emissions ($Lz^-$) and the
stimulated absorptions ($\bar n$) from the bath ($L^\dagger z^+$).
Notice also that for $T\rightarrow 0$, all the $\bar n_\omega$ 
tend to zero and (\ref{LQSDTGT0}) reduces to (\ref{LQSDT0}), as it 
should.

\subsection{Finite temperature and $L=L^\dagger$}
In the case of a selfadjoint coupling operator 
$L=L^\dagger\equiv K$, the finite temperature result can be simplified
considerably by introducing the sum process 
$z_t = z^-_t + z^+_t$ having zero mean and correlations
\beqa\label{qbmkernel}\nonumber
M[z_t^* z_s] & = &
\alpha^+(t,s) + \alpha^-(t,s) \equiv \alpha(t,s)
 = \sum_\omega \chi_\omega^2
[(2\bar n_\omega +1)\cos\omega(t-s)-i \sin\omega(t-s)]\\ 
M[z_t z_s] & = & 0.
\eeqa
Notice that $(2\bar n_\omega +1)=\coth\left(\frac{\hbar\omega}{2kT}\right)$ 
so that $\alpha(t,s)$ is nothing
but the well-known bath correlation kernel of the so-called quantum 
Brownian motion model \cite{qbm}.
In terms of this single process $z_t$, the linear non-Markovian
QSD equation at finite temperature (\ref{LQSDTGT0}) takes the simple
form of the zero-temperature equation (\ref{LQSDT0}) involving just
one noise $z_t$ 
\beq
\frac{d}{d t}\psi_t=-iH\psi_t + K\psi_t z_t - 
K\int_0^t \alpha(t,s)\frac{\delta\psi_t}{\delta z_s}ds,
\eeq
with the temperature dependent $\alpha(t,s)$ of (\ref{qbmkernel}).
For $K=q$ the position operator, this unravelling was first 
introduced in \cite{Strunz96b}, derived from the exact Feynman-Vernon 
path integral propagator of this model.

\end{appendix}

\newpage

\section*{Figure captions}

Fig 1a: Quantum trajectories of the non-Markovian QSD equation for
the `measurement'-like case
$H=\frac{\omega}{2}\sigma_z$, $L=\lambda\sigma_z$ and an exponentially
decaying bath correlation function 
$\alpha(t,s) = \frac{\gamma}{2}\exp(-\gamma|t-s|)$. We choose 
$\lambda^2 = 2\omega$, $\gamma = \omega$ and an initial state 
$|\psi_0\rangle = (1+2i)|\uparrow\rangle + (1+i)|\downarrow\rangle$.
Displayed is the expectation value $\langle \sigma_z \rangle$ of several
solutions of the non-Markovian QSD equation (solid lines) and
their ensemble average (dashed line).\\

Fig 1b: Same parameters as in Fig 1a: Here we compare the ensemble average 
of the 
Bloch vector using $10000$ quantum trajectories of non-Markovian QSD 
(solid lines), with the analytical result (dashed lines).\\

Fig 2a: Quantum trajectories of the non-Markovian QSD equation
for the dissipative case
$H=\frac{\omega}{2}\sigma_z$, $L=\lambda\sigma_-$ and an exponentially
decaying bath correlation function 
$\alpha(t,s) = \frac{\gamma}{2}\exp(-\gamma|t-s|-i\Omega(t-s))$. We 
choose 
$\lambda^2 = \omega$, $\gamma = \omega$ and resonance $\Omega=\omega$.
As initial state we use
$|\psi_0\rangle = 3|\uparrow\rangle + |\downarrow\rangle$.
Displayed is the expectation value $\langle \sigma_z \rangle$ of several
solutions of the non-Markovian QSD equation (solid lines) and
their ensemble average (dashed line).
At the finite time $\omega t_c=\frac{3}{2}\pi\approx 4.71$, 
all individual trajectories reach the ground state.\\

Fig 2b: Same parameters as in Fig 2a: 
Here we compare the ensemble average of the 
Bloch vector using $10000$ quantum trajectories of non-Markovian QSD 
(solid lines), with the analytical result (dashed lines).\\

Fig 3: Reversible decay of an initial Schr\"odinger cat state
$|\psi_0\rangle = |\alpha\rangle + |-\alpha\rangle$ with $\alpha=2$.
The contour plots show the $Q$-function of a non-Markovian quantum
trajectory of a harmonic oscillator ($\omega$), coupled to just a single
`environment' oscillator ($\Omega = 0.5\omega$), initially in its
ground state. The coupling strength
between the two oscillators is $0.1\omega$, and the time step between
two successive plots is $2.27/\omega$.\\

Fig 4: Shifting the `system-environment' boundary. First, we consider
the `spin - single oscillator' system with state $\psi_t(\xi)$, coupled
to a heat bath with noise $\xi_t$. Alternatively, we can consider the
`spin' only as the `system' $\phi_t(\xi,z)$, coupled to the
`single oscillator $+$ heat bath' environment (noises $\xi_t,z_t$).
In non-Markovian QSD, both descriptions are possible and lead
to the same reduced spin state.

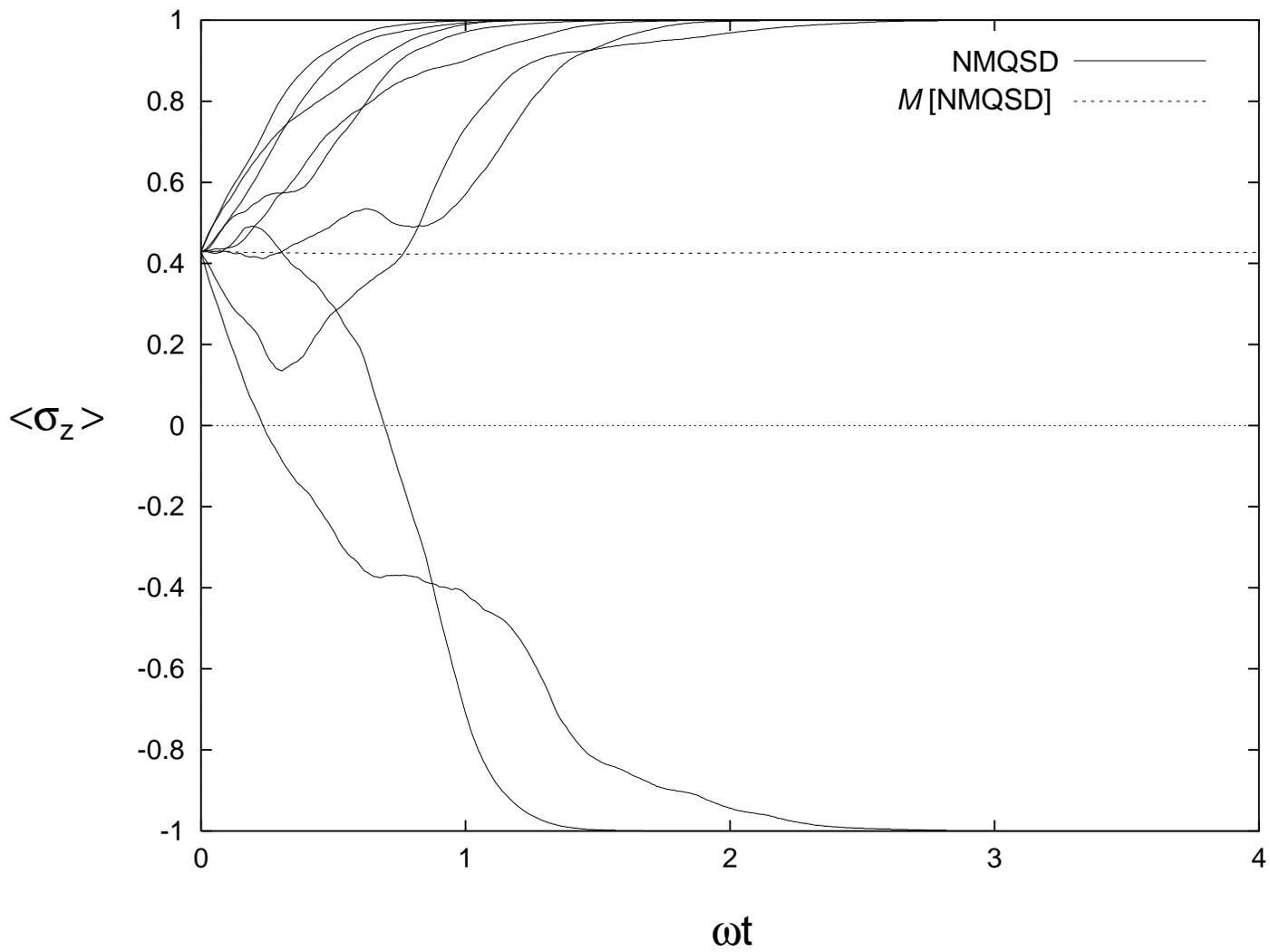
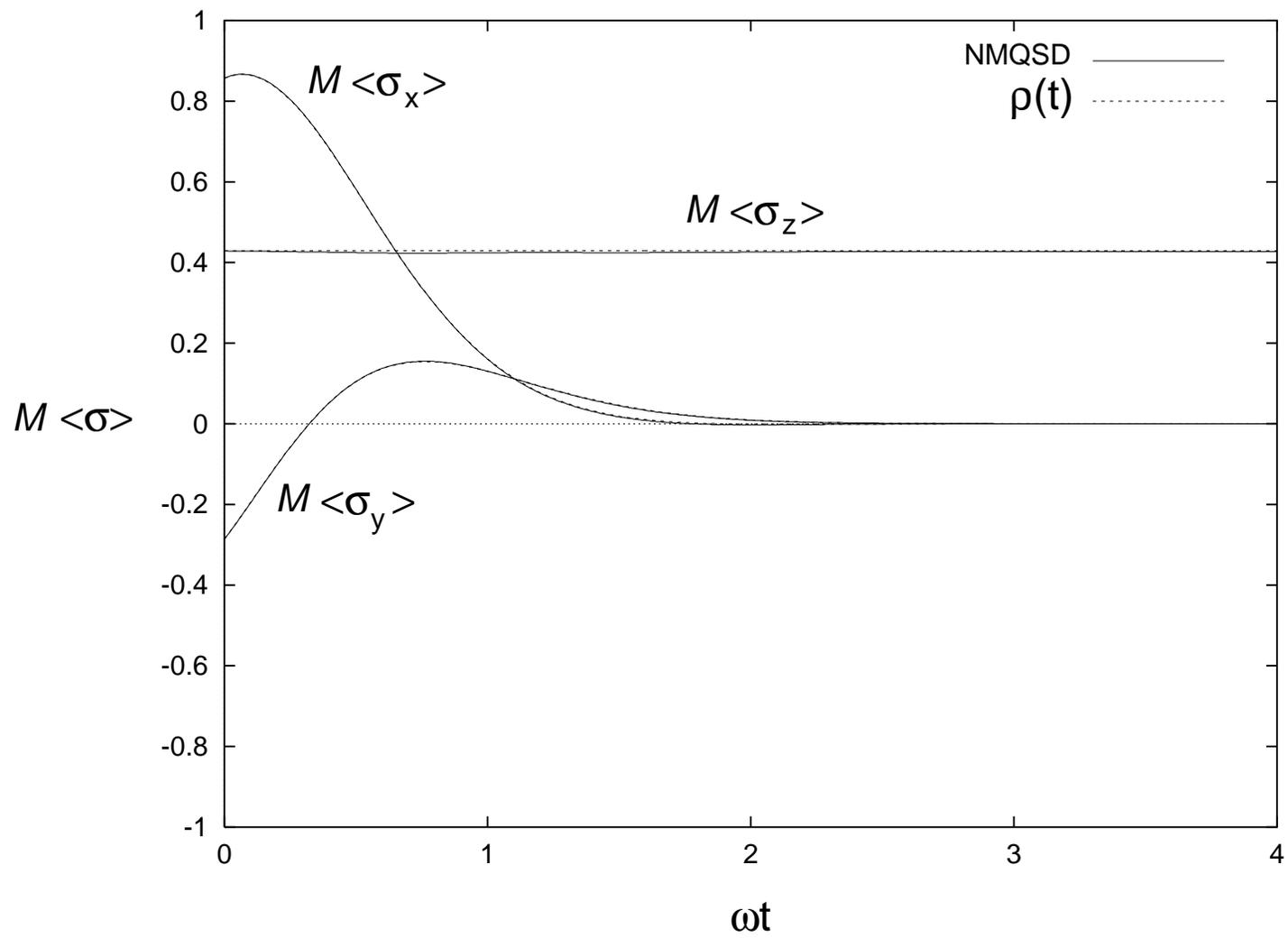
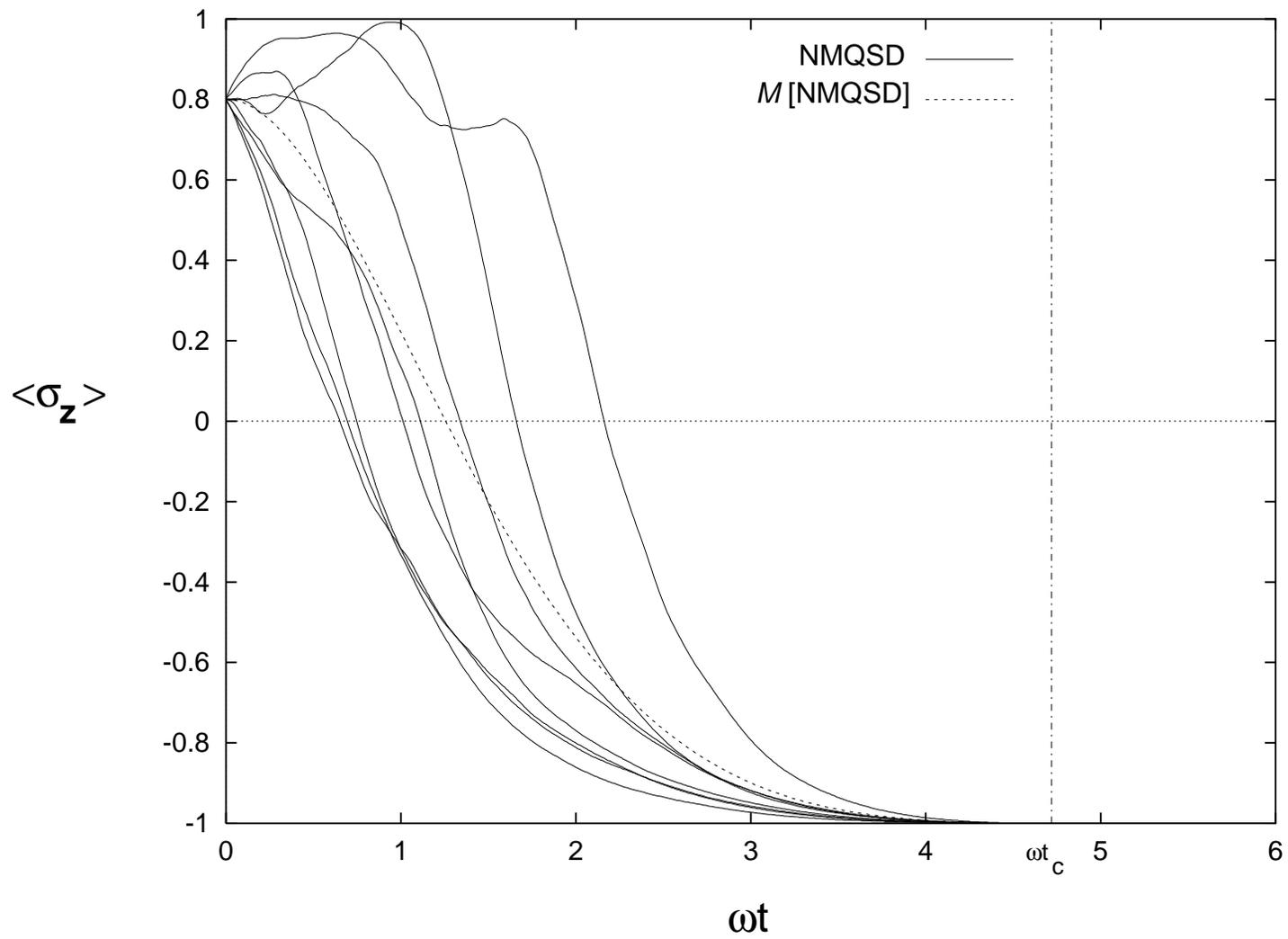
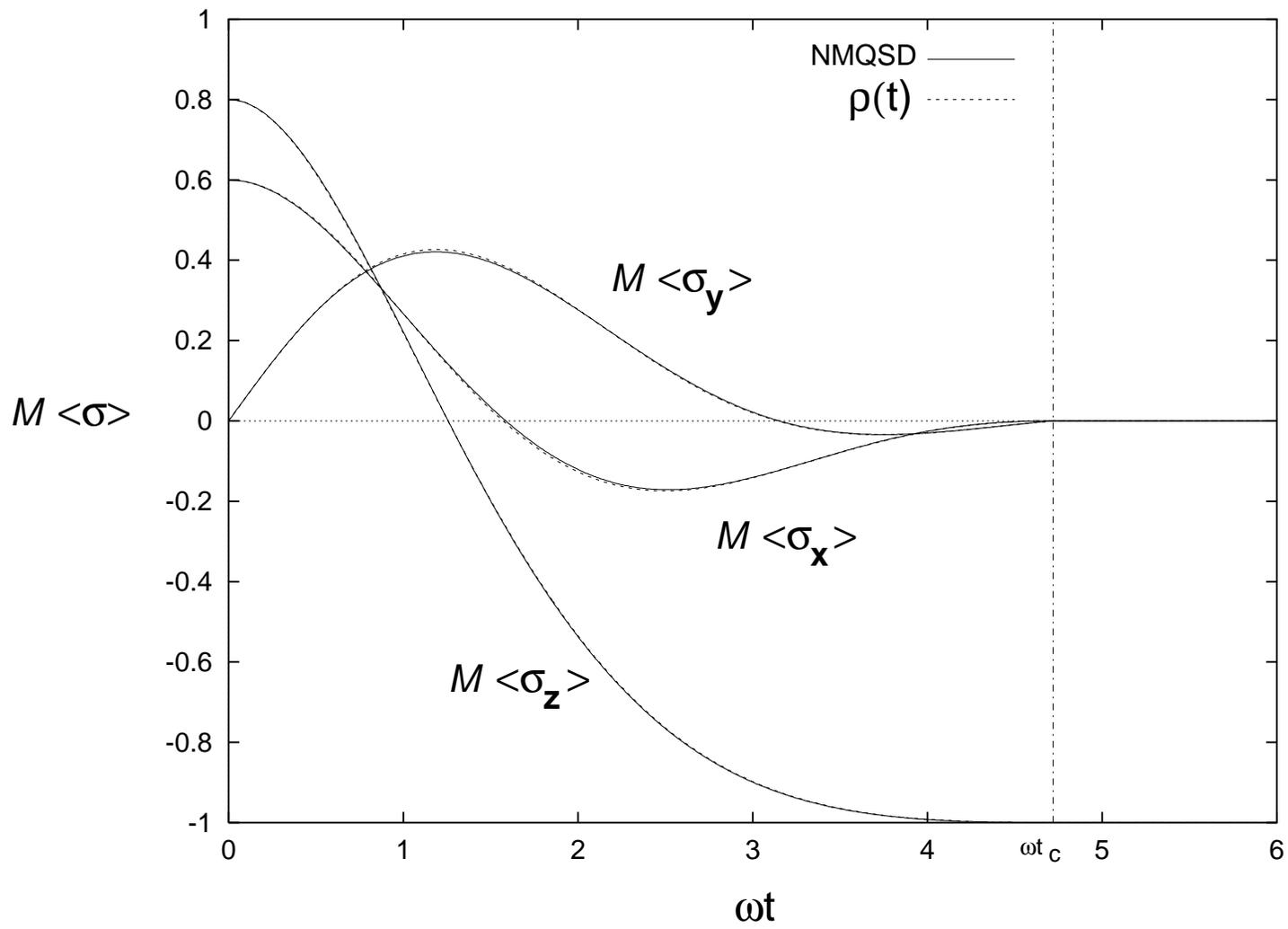
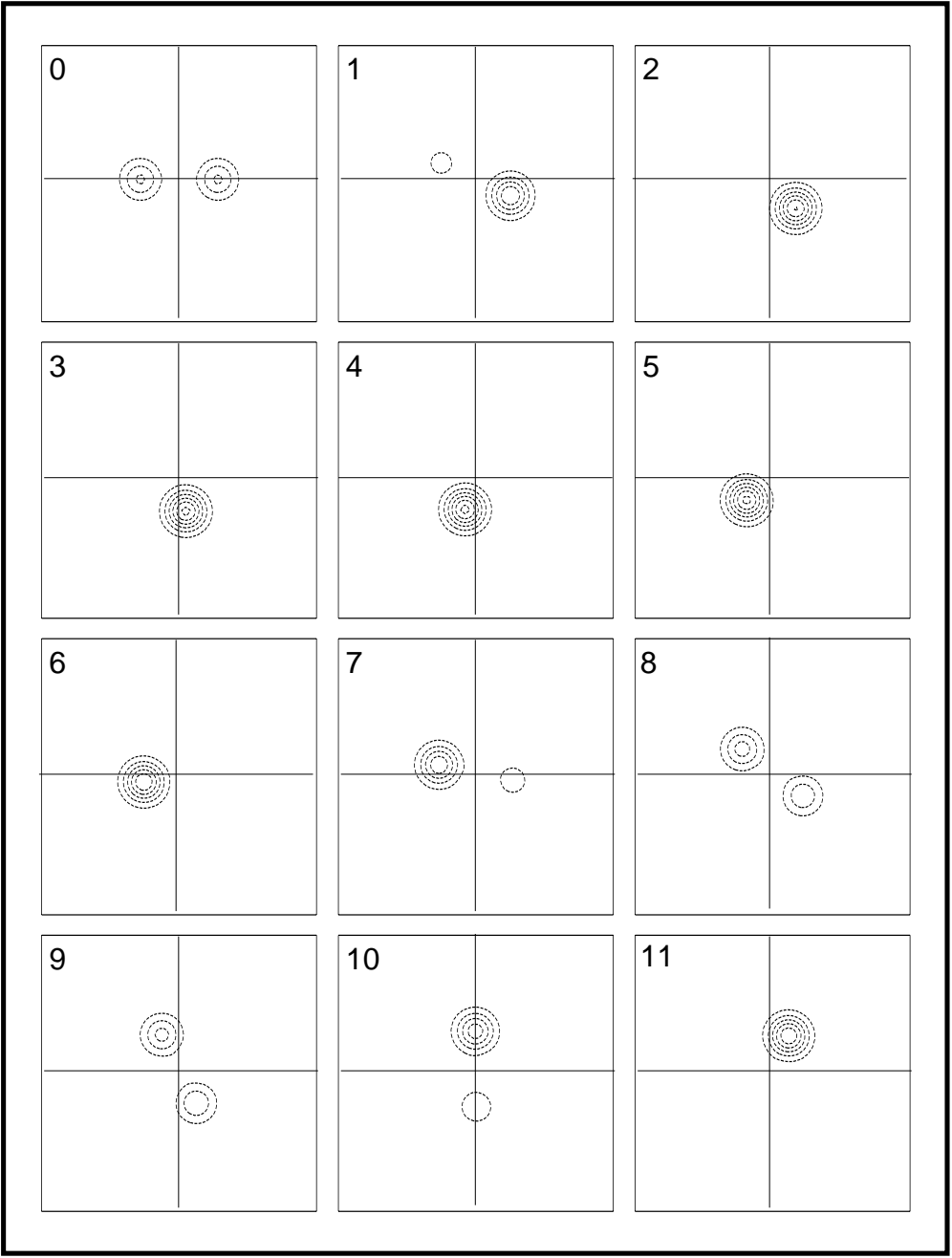
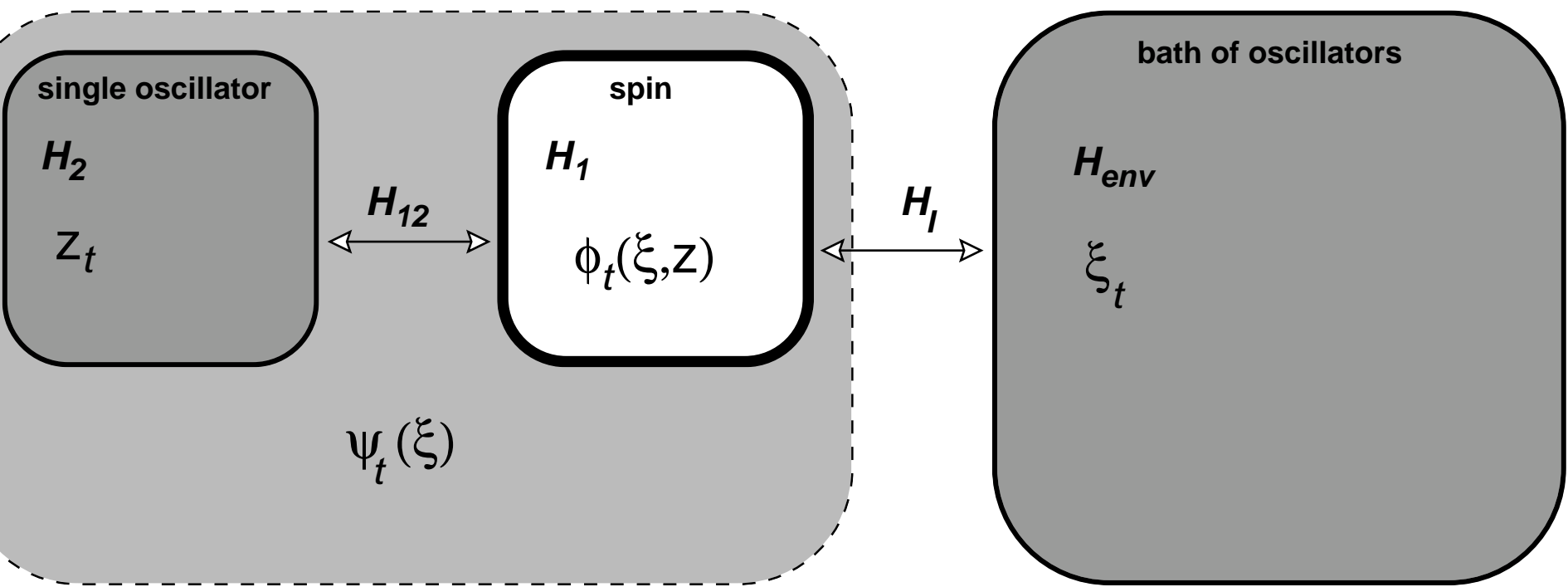
\end{document}